\documentclass[]{aa}
\usepackage{graphicx}

\begin{document}

\title{High-mass microquasars and low-latitude gamma-ray sources}

\author{V. Bosch-Ramon\inst{1} \and G. E. Romero\inst{2,3,}\footnote{Member of CONICET} 
\and J.~M. Paredes\inst{1}}

\institute{Departament d'Astronomia i Meteorologia, Universitat de Barcelona, Av. 
Diagonal 647, 08028 Barcelona, Spain; vbosch@am.ub.es, jmparedes@ub.edu.
\and Instituto Argentino de Radioastronom\'{\i}a, C.C.5,
(1894) Villa Elisa, Buenos Aires, Argentina; romero@iar.unlp.edu.ar.
\and Facultad de Ciencias Astron\'omicas y Geof\'{\i}sicas, UNLP, 
Paseo del Bosque, 1900 La Plata, Argentina.}

\offprints{V. Bosch-Ramon \\ \email{vbosch@am.ub.es}}

\abstract{Population studies of unidentified EGRET sources suggest that there exist 
at least three different populations of galactic gamma-ray sources. One of these 
populations is formed by young objects distributed along the galactic plane with 
a strong concentration toward the inner spiral arms of the Galaxy. Variability, 
spectral and correlation analysis indicate that this population is not homogeneous. 
In particular, there is a subgroup of sources that display clear variability in 
their gamma-ray fluxes on timescales from days to months. Following the proposal 
by Kaufman Bernad\'o et~al. (2002), we suggest that this group of sources might 
be high-mass microquasars, i.e. accreting black holes or neutron stars with 
relativistic jets and early-type stellar companions. We present detailed 
inhomogeneous models for the gamma-ray emission of these systems that 
include both external and synchrotron self-Compton interactions. We have 
included effects of interactions between the jet and all external photon 
fields to which it is exposed: companion star, accretion disk, and hot corona. 
We make broadband calculations to predict the spectral energy distribution of 
 the emission produced in the inner jet of these objects
up to GeV energies. The results and predictions can be 
tested by present and future gamma-ray instruments like INTEGRAL, AGILE, and 
GLAST. 
}

\maketitle

\section{Introduction}

Population studies of the unidentified gamma-ray sources detected by the EGRET instrument  of
the Compton Gamma-Ray Observatory (3rd EGRET catalog, Hartman et~al. \cite{Hartman99})
clearly suggest the existence of at least three  different groups of galactic sources
(Gehrels et~al. \cite{Gehrels00}; Grenier  \cite{Grenier01}, \cite{Grenier04}; Romero
\cite{Romero01}). There is a group  of weak and steady sources positionally correlated
with the Gould Belt (Grenier  \cite{Grenier95}, Gehrels et~al. \cite{Gehrels00}). These
sources are thought  to be nearby (100 -- 300 pc), with isotropic luminosities $\sim
10^{33}$ erg  s$^{-1}$. Following Romero et~al. (\cite{Romero04}), we will call these
sources  the Local Gamma-Ray Population (LGRP). There are $45\pm5$ sources in this
group.    

Another group of sources is concentrated along the Galactic Plane. They are 
well-correlated with star forming regions and HII regions, which is indicative 
of an association with young stellar objects (Romero et~al. \cite{Romero99}, 
Romero \cite{Romero01}). Log $N$ -- $\log S$ studies suggest that they are 
more abundant toward the inner spiral arms (Gehrels et~al. \cite{Gehrels00}, 
Bhattacharya et~al. \cite{Bhatta03}). These are bright sources (isotropic 
luminosities in the range $10^{34-36}$ erg s$^{-1}$), with an average photon 
spectral index  $\Gamma=2.18\pm 0.04$ ($F(E)\propto E^{-\Gamma}$). These 
sources, whose number is $\sim 45\pm 9$,
form the Gamma-Ray Population I (GRP I, Romero et~al. \cite{Romero04}). 

Finally, there is a third group of sources that are distributed forming a kind of  halo
around the galactic center. These sources have higher luminosities,  in the range
$10^{34-37}$ erg s$^{-1}$. They have soft spectra ($\Gamma\sim 2.5$)  and display strong
variability (Nolan et~al. \cite{Nolan03}). These sources should  be old, with ages from
hundreds of Myr to Gyr. About $45\pm 5$ EGRET detections fall  in this category (Grenier
\cite{Grenier01}). They form the Gamma-Ray Population II (GRP II).   

Among GRP I sources there is a subgroup that displays significant variability on 
timescales of weeks to months (Torres et~al. \cite{Torres01}, Nolan et~al.
\cite{Nolan03}).  Recently, Kaufman Bernad\'o et~al. (\cite{Kaufman02}) and Romero et~al.
(\cite{Romero04})  have suggested that this subgroup of GRP I sources might consist of
high-mass microquasars  (i.e. microquasars formed by a compact object and an early-type
stellar companion),  where the gamma-ray emission arises from interactions between
relativistic particles  in the jet and external photon fields, most notably the stellar UV
emission.   
Population studies by several authors (Grimm et~al. \cite{Grimm02}, Miyagi
\& Bhattacharya \cite{Miyagi04}) suggest that high-mass X-ray binaries (HMXB) are
distributed  in the Galaxy following the spiral structure, presenting a steeper $\log
N$ -- $\log S$ distribution than low-mass X-ray binaries (LMXB), which means 
that LMXB are more uniformly distributed in the Galaxy than HMXB. Since high-mass
microquasars form a subset of the whole set of  HMXB, they are also expected to be
distributed along the spiral arms, as is the case of GRP I sources.

In the  present paper we explore in more detail this hypothesis, presenting more
realistic  models for the gamma-ray emission. In particular, we will include effects of
the  interaction of the microquasar jet with the X-ray fields produced by the accretion 
disk and the hot corona that is thought to surround the compact object. We will  also
include synchrotron self-Compton emission, Klein-Nishina effects, and the back-reaction
of the different losses in the particle spectrum of the jet. We  will calculate, for some
representative sets of the parameters that characterize   high-mass microquasars, the
broadband spectral energy distribution (SED) of these objects. 

We will start discussing the main phenomenological properties of GRP I sources in order 
to clarify what we should expect from our models.  

\section{GRP I sources}

GRP I sources concentrate along the galactic plane and present a good spatial
correlation  with young stellar objects (Romero et~al. \cite{Romero99}). The variability
analysis of  these sources by Torres et~al. (\cite{Torres01}) clearly shows evidence for
the existence  of a subgroup with variable emission on timescales from weeks to months.
This is  corroborated by the recent results presented by Nolan et~al. (\cite{Nolan03}),
which  are based on a maximum likelihood re-analysis of the EGRET data. These authors
identify 17  variable sources within $\mid 6^{\circ}\mid$ from the galactic plane. These
sources are clumped  within $\mid 55^{\circ}\mid$ of the galactic center.

A $\log N - \log S$ analysis for all GRP I sources yields a distribution that is 
consistent with a power-law with index $\beta\sim 3.1$ (Bhattacharya et~al. 
\cite{Bhatta03}). 
This is far steeper than what is expected for a population uniformly 
distributed along the galactic disk. For instance, for pulsars detected at 400 MHz the 
slope is $\beta\sim1.05$. The unidentified gamma-ray sources, on the contrary, 
seem to be concentrated mainly in the inner spiral arms. To find possible 
further evidence for different populations among GRP I sources, we have implemented a 
$\log N - \log S$ analysis of both variable and non-variable low-latitude sources.

\begin{figure}
\resizebox{\hsize}{!}{\includegraphics{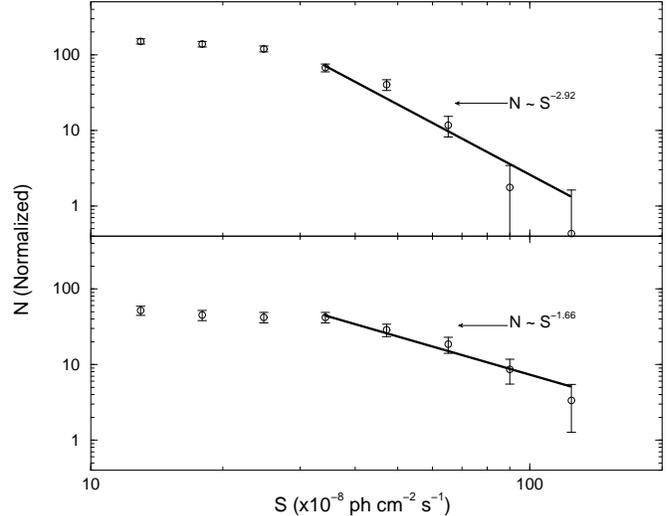}}
\caption{Log $N - \log S$ plot for gamma-ray sources within $|b|<6^{\circ}$.
\textbf{a)} Upper panel: For non variable gamma-ray sources.
\textbf{b)} Lower panel: For variable gamma-ray sources.}
\label{Fig1}
\end{figure}

First we have considered the 17 variable sources listed by Nolan et~al. (\cite{Nolan03}). 
To take into account systematic effects introduced by different exposure and 
background resulting in non-uniform detectability, we have adopted the procedure 
described by Reimer (\cite{Reimer01}). The obtained $\log N - \log S$ plot is shown 
in Fig. \ref{Fig1}, lower panel. The normalized distribution can be fitted by a power-law 
$N(S)\propto S^{-\beta}$, with  $\beta=1.66\pm0.31$, significantly harder than for 
the entire sample.
If we now consider those sources that classify as non-variable 
or dubious cases, we get the $\log N - \log S$ plot shown  
also in Fig. \ref{Fig1}, upper panel. 
In this case the distribution can be fitted by a power-law with index $\beta=2.92\pm0.36$.       

The average spectral index is also different for both samples: in the case of the variable 
sources we have $\left\langle \Gamma \right\rangle=2.04\pm 0.03$, whereas for the remaining 
sources $\left\langle \Gamma \right\rangle=2.16\pm 0.01$. All this suggests that there 
are two different groups of sources, one formed by steady sources concentrated toward 
the inner spiral arms, and a second group with variable sources and a wider distribution 
along the galactic plane, although not as wide as that of radio pulsars (Bhattacharya 
et~al. \cite{Bhatta03}).

Microquasars appear to be good candidates for compact and variable sources in
the  galactic plane. Since they can have very large proper motions (e.g. Rib\'o
et~al.  \cite{Ribo02}, Mirabel et~al. \cite{Mirabel04}), their distribution
along the plane should be broader than that presented  by supernova remnants
and molecular clouds, which can be traced by star-forming regions  and OB
associations. Their spread, however, is limited by the lifespan of the
companion  massive star, and hence it is not as extended as that of radio
pulsars. It is worth noting that Miyagi \& Bhattacharya (\cite{Miyagi04})
obtained for HMXB  a $\beta$ of about 1.9. In the next  section we will
discuss the potential of microquasars to be gamma-ray sources.    

\section{Gamma-ray emission from microquasars}

Microquasars are variable sources at all wavelengths where they have been detected.  At
radio frequencies they present non-thermal jets and in some cases superluminal 
components (Mirabel \& Rodr{\'{\i}}guez \cite{Mirabel&Rodriguez99}). In the case of
high-mass microquasars, with which we are concerned here, the emission at optical and  UV
energies is dominated by the companion star, where it can reach luminosities of $\sim
10^{39}$ erg s$^{-1}$. Accretion  onto the compact object results in the formation of an
accretion disk with typical  temperatures of a few keV and thermal luminosities that can
reach $\sim 10^{37}$ erg  s$^{-1}$. In the low-hard state, when the jets appear to be
stronger, microquasars usually  present a hard X-ray component that can be represented by
a power-law plus an  exponential cutoff at a few hundred keV. This component might be
produced by a hot  corona or ADAF region around the central object (e.g. Poutanen
\cite{Poutanen98}, McClintock \& Remillard \cite{McClin04})  although some authors have
argued for a purely non-thermal origin in the jets (Markoff et~al. \cite{Markoff01},
\cite{Markoff03}; Georganopoulos  et~al. \cite{Georganopoulos02}).  Beyond the problem of
the exact nature of the different contributions to the hard X-rays,  it is clear that
very relativistic electrons are present in the jets of microquasars.  Since these
electrons must traverse several photons fields, inverse Compton interactions  are
unavoidable. Such interactions can produce high-energy gamma-rays with a luminosity  that
will depend on the particular physical parameters that characterize the source. 

In recent years, several authors have studied models for leptonic gamma-ray 
production in microquasars (Atoyan \& Aharonian \cite{Atoyan&Aharonian99}, 
Paredes et~al. \cite{Paredes00}, Paredes et~al. \cite{Paredes02}, Kaufman Bernad\'o et~al. \cite{Kaufman02},
Romero et~al. \cite{Romero02}, Bosch-Ramon \& Paredes \cite{Bosch04}). 
Most of these models consider steady and compact jets. Such jets are usually
associated with the low-hard X-ray state of the sources. However, some
microquasars present persistent radio jets and moderate X-ray emission. This is the
case of the two already known microquasars that have been proposed to be
counterparts of EGRET sources: LS 5039 (Paredes et al. \cite{Paredes00},\cite{Paredes02}) and
LS I +61 303 (Massi et al. \cite{Massi04}). The fact that a new microquasar
candidate with similar X-ray properties has been recently discovered by Combi
et al. (\cite{Combi04}) within the EGRET location error box of 3EG J1639-4702,
lends additional support to the idea that perhaps high-energy gamma-ray emitting
microquasars are not typical X-ray binary systems from the point of view of
their X-ray behavior.

In the present paper we will consider a microquasar with a continuous, persistent and inhomogeneous jet endowed 
with a structure similar to that adopted by Romero et~al. (\cite{Romero03}). 
In our case, however, we will focus on the leptonic content.
The jet is 
ejected along the rotation axis of the compact object, which for simplicity 
is assumed 
to be perpendicular to the orbital plane
\footnote{Misaligned jets are probably a common 
case but our results will not be significantly affected by a small inclination 
angle. See Maccarone (\cite{Maccarone02}).}.

The relativistic jet flow will move along the $z$ axis with a bulk Lorentz factor 
$\Gamma_{\rm jet}$ and a constant velocity $\beta=v/c$. Hadrons, 
although carrying most of the kinetic power, will play no radiative role in our 
model (for hadronic gamma-ray emission see the paper by Romero et~al. 
\cite{Romero03}). We shall allow the jet to expand laterally, in such a way that 
the radius $R$ at a distance $z$ from the compact object will be given by 
$R(z)=\xi z^{\varepsilon}$, with $\varepsilon\leq 1$ and $z_0\leq z\leq z_{\rm max}$. 
For $\epsilon=1$ we have a conical jet. 

The relativistic electrons in the jet are assumed to have an energy distribution given 
by a power-law, as inferred from the observed synchrotron emission: 
\begin{equation}
N(z,\gamma_{\rm e})=k(z)\gamma_{\rm e}^{-p} 
\label{eq:distfunc}
\end{equation}
\begin{equation}
k(z)=k(z_0)\left(\frac{R_0}{R}\right)^2=k(z_0)\left(\frac{z_0}{z}\right)^{2\varepsilon}.
\label{eq:constevol}
\end{equation}
Here, $p$ is the power-law index of the electron energy distribution and $R_0=R(z_0)$ is the radius of 
the jet at the injection point. 
$\gamma_{\rm e}$ is the electron Lorentz factor and $N(z,\gamma_{\rm e})$ is the 
number of electrons of a given energy per unit
of volume at a given distance from the compact object. The electron distribution does not
depend on the direction of the particle velocity. 

Another important parameter associated with the electron energy distribution is the 
maximum Lorentz factor, $\gamma_{\rm emax}$. Its evolution is described by:
\begin{equation}
\gamma_{\rm e max(z)}=\gamma_{\rm e
max}(z_0)\left(\frac{z_0}{z}\right)^{e\varepsilon}.
\label{eq:gmaxevol}
\end{equation}
The parameter $e$ is introduced to take into account loss or gain
energy processes other than adiabatic losses, which are already counted through
$\varepsilon$, without a deep description of the involved physical processes (see, i.e.,
Ghisellini et~al. \cite{Ghisellini85}). Both $\varepsilon$ and $e$ cover our lack of
knowledge of what happens to/within the jet. In the present work, we will set $e$ to 1
from now on, to simplify the number of free parameters, assuming as a first approximation
a jet controlled by adiabatic evolution. We will come back to this issue in
subsect.~\ref{elenev}, analysing more carefully this question.

Depending on the geometrical nature of the flow and assuming
adiabatic expansion, the magnetic field changes with $z$ as: 
\begin{equation}
B(z)=B(z_0)\left(\frac{R_0}{R}\right)=B(z_0)\left(\frac{z_0}{z}\right)^{\varepsilon}.
\label{eq:Bevol}
\end{equation}
Then, the synchrotron radiation density can be estimated as:
\begin{eqnarray}
U_{\rm syn,\, \nu}(z)&\approx&\frac{s_{\nu}(z)}{c}\tau_{\nu}(z)\nonumber\\
&=&
\frac{C_5}{cC_6}B(z)^{-1/2}\left(\frac{\nu}{2C_1}\right)^{5/2}
\left(\frac{\nu}{\nu_1(z)}\right)^{-(p+4)/2}, 
\label{eq:syncemis}
\end{eqnarray}
where $s_{\nu}$ is the source function of the synchrotron emission 
from an isotropic particle 
distribution and $\tau_{\nu}$ is the synchrotron optical depth of the jet. The explicit 
expressions for both $s_{\nu}$ and $\tau_{\nu}$ can be found in Pacholczyk 
(\cite{Pacholczyk70}). We use the formulae in the optically thin case because the 
synchrotron emission mainly comes from frequencies beyond the self-absorption 
frequency, which is given by
\begin{equation} 
\nu_1(z)\approx2C_1[R(z)C_6k(z)(m_{\rm
e}c^2)^{(p-1)}]^{\frac{2}{p+4}} B(z)^{\frac{p+2}{p+4}},
\end{equation}
 where $C_1$, $C_5$ and
$C_6$ are numerical constants given in Table~\ref{constants}. This is the local
approximation to the synchrotron radiation field (Ghisellini et~al. \cite{Ghisellini85}).

\begin{table}
\begin{center}
\caption[]{Constants in synchrotron formulae.}
\begin{tabular}{l c c c c c}
\noalign{\smallskip} 
\hline 
\hline 
\noalign{\smallskip} Parameter (symbol) & Value 
\cr \noalign{\smallskip} \hline \noalign{\smallskip} 
$C_1$ & $6.27\times10^{18}$~cgs~units
\cr $C_5(p=2)$ & $1.37\times10^{-23}$~cgs~units
\cr $C_6(p=2)$ & $8.61\times10^{-41}$~cgs~units
\cr \noalign{\smallskip} \hline
\end{tabular}
\label{constants}
\end{center}
\end{table}

The total radiation field to which the leptons are exposed in the jet also will have 
a contribution from external sources. These contributions can be modeled as two black body 
components, one peaked at UV energies (the companion star field) and other at energies 
$kT\sim 1$ keV (the inner accretion disk field), plus a power-law with an exponential 
cutoff at $kT\sim 150$ keV (the corona). With the exception of the disk, 
these contributions 
are assumed to be isotropic (Romero et~al. \cite{Romero02}, Georganopoulos  et~al. 
\cite{Georganopoulos02}).

Once $N(z,\gamma_{\rm e})$ and the total energy density in seed photons 
($U_{\rm total}(\epsilon_0,z)$) have been determined, the IC interaction between them can be studied
calculating the IC spectral energy distribution per energy unit. The cross section for both the
Thomson and the Klein-Nishina regimes has been approximated by Blumenthal \& Gould
(\cite{Blumenthal&Gould70}):
\begin{equation}
\frac{d\sigma(x,\epsilon_0,\gamma_{\rm e})}{d\epsilon}=\frac{3\sigma_{\rm
T}c}{4 \epsilon_0\gamma_{\rm e}^2}~f(x),
\label{eq:crosssection}
\end{equation}
where
\begin{eqnarray}
f(x)&=&\left[2x \ln x+x+1-2x^2+\frac{(4\epsilon_0\gamma_{\rm e}
x)^2}{2(1+4\epsilon_0\gamma_{\rm e} x)}(1-x)\right]\nonumber \\ && \times \; P(1/4\gamma_{\rm
e}^2,~1,~x),
\end{eqnarray}
\begin{equation}
x=\frac{\epsilon}{4\epsilon_0\gamma_{\rm
e}^2(1-\frac{\epsilon}{\gamma_{\rm e}})},
\end{equation}
$\epsilon_0$ and $\epsilon$ are the energies of the incoming and the outgoing photons,
respectively, $\sigma_{\rm T}$ is the Thomson cross section, and
\begin{equation}
P(1/4\gamma_{\rm e}^2,~1,~x)=1,~~$for$~~1/4\gamma_{\rm e}^2 \le x \le
1,
\end{equation}
and 0 otherwise.

The spectral energy distribution for the optically thin case in the
jet's reference frame is:
\begin{eqnarray}
\epsilon L_{\epsilon}&=&\epsilon
\int^{z_{\rm max}}_{z_{\rm min}}
\int^{\epsilon_{\rm 0 max}(z)}_{\epsilon_{\rm 0 min}(z)}
\int^{\gamma_{\rm emax}(z)}_{\gamma_{\rm emin}(z)}
\Sigma(z)U_{\rm tot}(\epsilon_0,z) \nonumber \\&& 
\times  N(\gamma_{\rm e},z) \frac{d\sigma(x,\epsilon_0,\gamma_{\rm e})}{d\epsilon}
\frac{\epsilon}{\epsilon_0}d\gamma 
d\epsilon_0 
dz,
\label{eq:Lic}
\end{eqnarray}
where $\Sigma(z)$ is the surface of a perpendicular jet slice located at $z$.

Notice that the external fields contributing to $U_{\rm tot}(\epsilon_0,z)$ 
should be transformed to the co-moving frame. Detailed expressions for such 
transformations are given by Dermer \& Schlickeiser (\cite{Dermer&Sch02}). 
In the observer's reference frame we have:
\begin{eqnarray}
\epsilon' L'_{\epsilon'}&=& D^{2+p} \epsilon' L_{\epsilon'}.
\label{eq:LicSRobs}
\end{eqnarray}
The integration is performed in the co-moving system and then the result is transformed
to the  observer's frame, hence the factor $D^{2+p}$, which is the Doppler boosting for
a continuous  jet. The energy of the scattered photon in the jet's reference frame
($\epsilon$)  is boosted to $\epsilon'=D\epsilon$. The Doppler factor $D$ for the 
approaching jet is given by  
\begin{equation}
\frac{1}{\Gamma_{\rm jet}(1-\beta\cos\theta)},
\end{equation}
where $\beta$ is the velocity of the jet in speed of light units and $\theta$ is the
angle  between the jet and the line of sight. We note that 
$P(1/4\gamma_{\rm e}^2,~1,~x)$ restricts the range of $x$ to physical values,  where
$\epsilon$ cannot be lower than $\epsilon_0$ ($x=1/4\gamma_{\rm e}^2$) or  larger than
$\epsilon_{\rm max}$ ($x=1$). 

In the case of the IC interactions with disk photons, a factor $(1-\cos
\theta)^{(p+1)/2}$ must  be introduced in Eq.(~\ref{eq:Lic}) to take into
account the fact that the photons come from behind the jet (Dermer et~al.
\cite{Dermer92}).  

To make any calculation of the IC emission of a given microquasar, we have to specify 
first the jet power in leptons and hence the constant $k$ in Eq. (\ref{eq:distfunc}). In this work we 
shall adopt the disk/jet coupling hypothesis formulated by Falcke \& Biermann  
(\cite{Falcke&Biermann95}, \cite{Falcke&Biermann99}), i.e. the jet power is proportional 
to the accretion rate:
\begin{equation}
L^{\rm lep}_{\rm jet}=\pi R^2
c \Gamma_{\rm jet}\beta \int m_{\rm e} c^2 k(z_0) \gamma^{1-p}d\gamma=q_{\rm e}L_{\rm acc}.
\end{equation}
Here, $L_{\rm acc}=\dot{M_{\rm acc}}c^2$ is the accretion power onto the compact object 
and $q_{\rm e}$ 
is a number $<1$. The total jet power is 
$L_{\rm jet}=q_{\rm tot}L_{\rm acc}=(q_{\rm hadrons}+q_{\rm e})L_{\rm acc}$. If the 
hadrons are also relativistic, $q_{\rm hadrons}\sim q_{\rm tot} >>q_{\rm e}$. 
Typically, $q_{\rm tot}\sim 10^{-1}-10^{-3}$ and then $q_{\rm e}$ is, in the kind of 
jet we are considering, in the range $\sim 10^{-3}-10^{-5}$. 

\section{Specific assumptions}

We make a number of specific assumptions regarding the characterization of the high-mass 
microquasar model adopted in the calculations of the SED. In particular, we will consider  a
system where the compact object is a black hole of 10~$M_{\odot}$ that accretes 
$10^{-8}$~$M_{\odot}$~yr$^{-1}$. The gravitational radius $R_{\rm g}$ for such an object is 
$\sim 1.5\times 10^6$~cm. The companion star has a radius of $R_*=15~R_{\odot}$ and a
bolometric luminosity  of $\sim 5\times10^{38}$~erg~s$^{-1}$. Its radiation field is well
represented by a black body with  $kT\sim 10$~eV. The orbital radius is $R_{\rm orb}\sim 
45~R_{\odot}$. 

Around the black hole there is a thermal disk with typical temperatures given by $kT\sim 1$~keV. The
total disk luminosity is  $\sim10^{37}$~erg~s$^{-1}$. The hot corona above the central disk is
represented by a power law with a hard photon  index $1.6$ and an exponential cutoff at $\sim150$~keV.  

Regarding the 
jet, we assume it is injected at $z_0\sim 50~R_{\rm g}$. Its initial 
radius is $R_0\sim 0.1~z_0$ and its lateral expansion is characterized by a coefficient 
$\varepsilon=1$. The electron energy distribution has a power law index $p=2$. 
The high-energy cutoff is assumed at $\gamma_{\rm emax}(z_0)=10^4$, 
whereas the minimum Lorentz factor is $\gamma_{\rm emin}\sim 1$.

The magnetic field outside the coronal region, $B(z_0)$, is unknown. Most
models for jet production require high magnetic fields, but these fields are
usually attached to the inner accretion disk. Here we will assume a set of
different possible values, in the range $0.1-200$ G. These values are below
equipartition, which could be reached further in the jet (e.g. in the radio
emitting region).

All these assumptions are summarized in Table \ref{paramvalues}.

\begin{table*}
\begin{center}
\caption[]{Fixed parameters in the models.}
\begin{tabular}{l c c c c c}
\noalign{\smallskip} \hline 
\hline 
\noalign{\smallskip} Parameter (symbol) & Value 
\cr \noalign{\smallskip} \hline \noalign{\smallskip} 
Black hole mass ($M_{\rm bh}$) & $10 M_{\odot}$  
\cr Gravitational radius ($R_{\rm g}$) & $1.48\times10^6$~cm  
\cr Accretion luminosity ($L_{\rm acc}$) & $10^{-8}~M_{\odot} c^2$~year$^{-1}$
\cr Stellar radius ($R_{*}$) & $15~R_{\odot}$  
\cr Stellar bolometric luminosity ($L_{*}$) & 5$\times10^{38}$~erg~s$^{-1}$
\cr Viewing angle to jet's axis $\theta$ & $10^{\circ}$
\cr Distance from jet's apex to the compact object ($z_0$) & $50~R_{\rm g}$ 
\cr Initial jet radius ($R_0$) & 0.1$z_0$
\cr Orbital radius ($R_{\rm orb}$) & $45~R_{\odot}$
\cr Peak energy of the disk ($kT_{\rm disk}$) & 1~keV
\cr Peak energy of the corona & 150~keV
\cr Peak energy of the star ($kT_{\rm star}$) & 10~eV
\cr Expansion coefficient of the jet ($\varepsilon$) & 1
\cr Minimum Lorentz factor for electrons in the jet (jet frame) ($\gamma_{\rm emin}$) & $\sim 1$
\cr Maximum Lorentz factor for electrons in the jet (jet frame) ($\gamma_{\rm emax}$) & $10^4$
\cr Electron energy distribution power-law index ($p$) & 2
\cr Photon index for the corona ($\Gamma_{\rm cor}$) & 1.6
\cr Total disk luminosity ($L_{\rm disk}$) & $10^{37}$~erg~s$^{-1}$
\cr \noalign{\smallskip} \hline
\end{tabular}
\label{paramvalues}
\end{center}
\end{table*}

\section{Calculations} \label{exploring}

We have calculated the high-energy SED for a variety of cases using the typical values 
of the parameters given in Table~\ref{paramvalues}, and different combinations of magnetic 
field $B(z_0)$, luminosity of the corona ($L_{\rm cor}$), bulk Lorentz factor 
$\Gamma_{\rm jet}$ and disk/jet coupling parameter $q_{\rm e}$. In Table~\ref{frparam} we give the 
range of these parameters adopted in our calculations. 

\begin{figure}
\resizebox{\hsize}{!}{\includegraphics{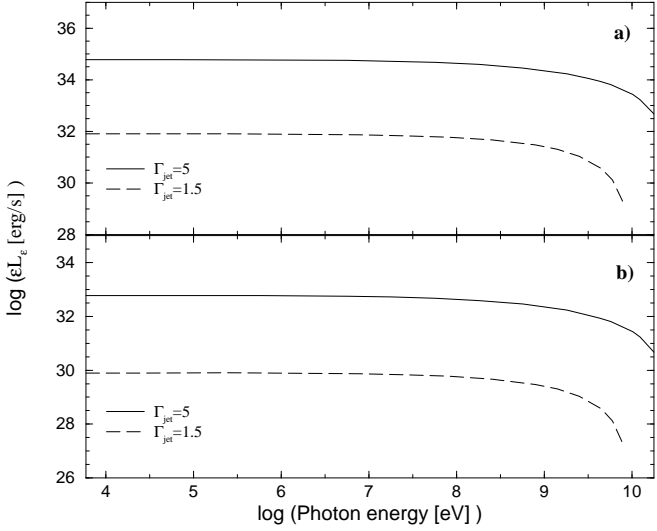}}
\caption{High-energy SED from IC up-scattering of stellar photons. 
\textbf{a)} Models with $q_{\rm e}=10^{-3}$: for $\Gamma_{\rm jet}=$1.5 and 5.
 \textbf{b)} Models with $q_{\rm e}=10^{-5}$: for $\Gamma_{\rm jet}=$1.5 and 5.}
\label{fig3}
\end{figure}

\begin{figure}
\resizebox{\hsize}{!}{\includegraphics{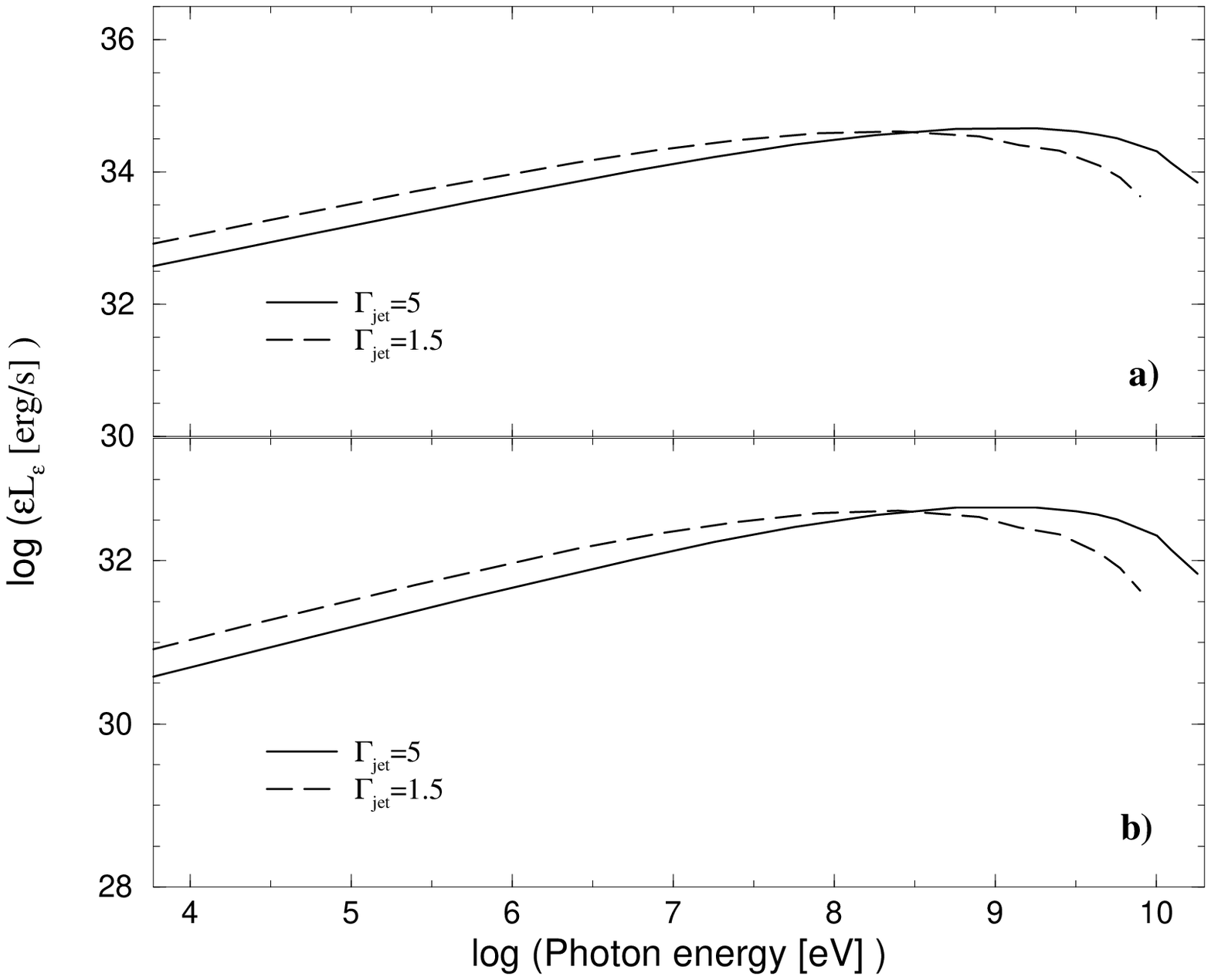}}
\caption{High-energy SED from IC up-scattering of disk photons. 
\textbf{a)} Models with $q_{\rm e}=10^{-3}$: for $\Gamma_{\rm jet}=$1.5 and 5.
 \textbf{b)} Models with $q_{\rm e}=10^{-5}$: for $\Gamma_{\rm jet}=$1.5 and 5.}
\label{fig4}
\end{figure}

\begin{figure}
\resizebox{\hsize}{!}{\includegraphics{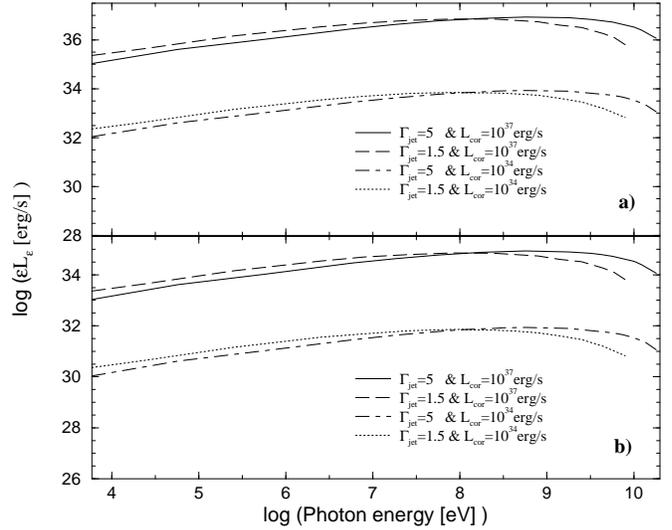}}
\caption{High-energy SED from IC up-scattering of corona photons. 
\textbf{a)} Models with $q_{\rm e}=10^{-3}$: for $\Gamma_{\rm jet}=$1.5, 5; and 
$L_{\rm cor}=10^{34}$, $10^{37}$~erg~s$^{-1}$.
\textbf{b)} Models with $q_{\rm e}=10^{-5}$: for $\Gamma_{\rm jet}=$1.5, 5; and
$L_{\rm cor}=10^{34}$, $10^{37}$~erg~s$^{-1}$.}
\label{fig5}
\end{figure}

\begin{figure}
\resizebox{\hsize}{!}{\includegraphics{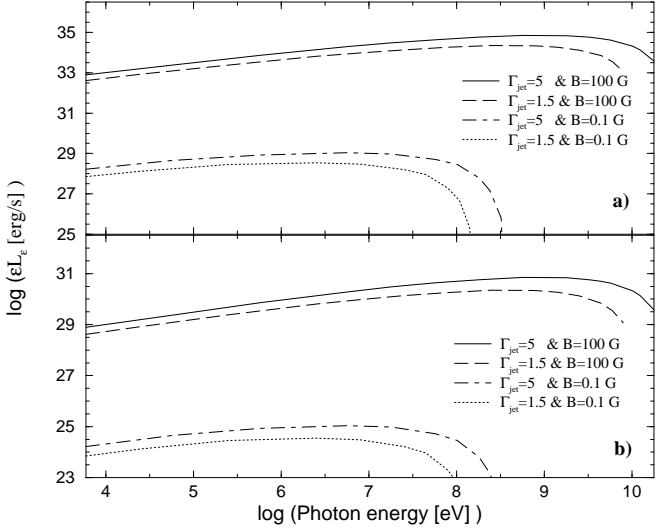}}
\caption{High-energy SED from IC up-scattering of synchrotron photons. 
\textbf{a)} Models with $q_{\rm e}=10^{-3}$: for $\Gamma_{\rm jet}=$1.5, 5; and 
$B(z_0)$=0.1, 100~G.
\textbf{b)} Models with $q_{\rm e}=10^{-5}$: for $\Gamma_{\rm jet}=$1.5, 5; and
$B(z_0)$=0.1, 100~G.}
\label{fig6}
\end{figure}

In Figs. \ref{fig3} to \ref{fig6}, we show some representative results for the SED produced 
by the IC up-scattering of stellar, disk, corona, and synchrotron photons, respectively. 
Curves obtained from different combinations of the parameters are shown in each panel. 

\begin{figure}
\resizebox{\hsize}{!}{\includegraphics{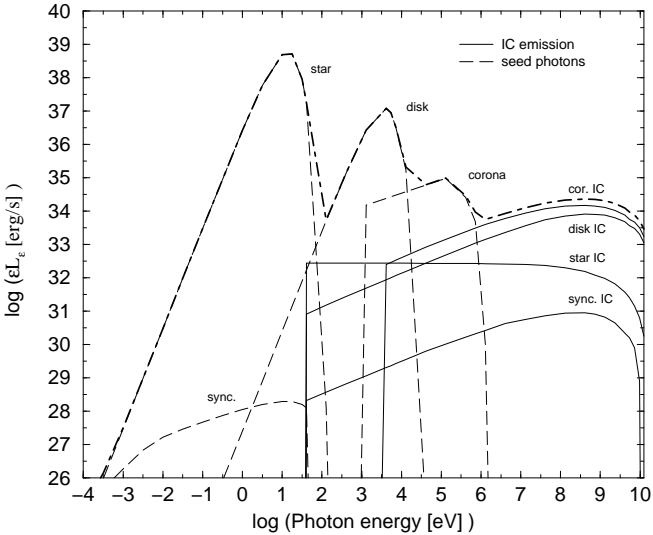}}
\caption{Complete SED for a model with $q_{\rm e}=10^{-4}$, $B(z_0)$=10~G, 
$L_{\rm cor}=10^{35}$~erg~s$^{-1}$, $\gamma_{\rm emax}(z_0)=10^4$, $\Gamma_{\rm jet}$=2.5, 
and a viewing angle of $10^{\circ}$. This is the representative case.} 
\label{ideal}
\end{figure}

\begin{figure}
\resizebox{\hsize}{!}{\includegraphics{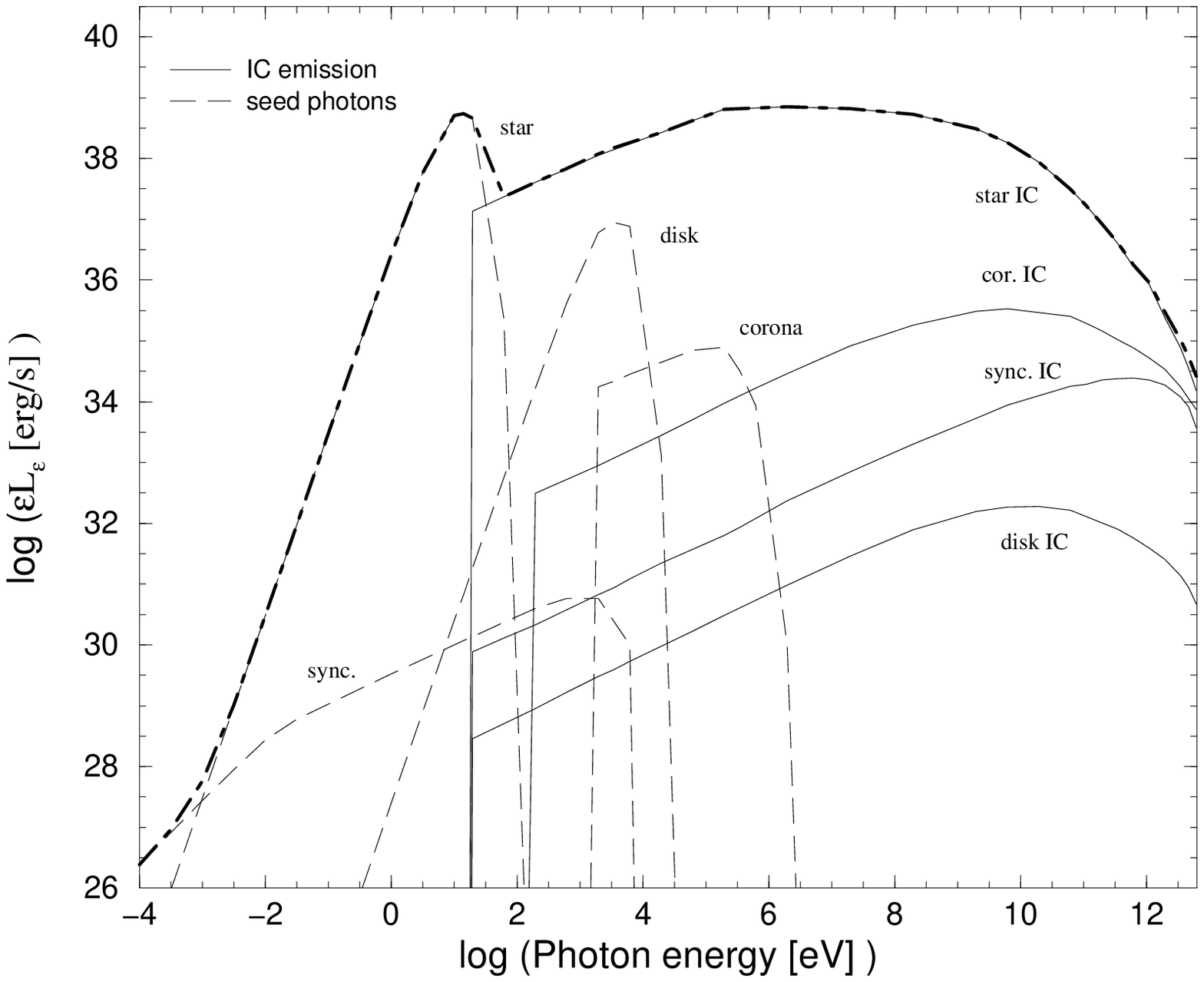}}
\caption{SED for a model with $q_{\rm e}=10^{-4}$, $B(z_0)$=10~G, $L_{\rm
cor}=10^{35}$~erg~s$^{-1}$, $\gamma_{\rm emax}(z_0)=10^6$, $\Gamma_{\rm jet}$=10, 
and a viewing angle of $1^{\circ}$. This is the extreme case.} 
\label{extreme}
\end{figure}

In Fig.~\ref{ideal}, we show the whole SED of a representative case, with the parameters
listed in Table~\ref{tideal1}. We plot here the spectra for all contributions  to the
emission: the seed photon sources and the IC components. In Fig.~\ref{extreme} we show a
more extreme case, for a source with a high bulk Lorentz factor $\Gamma_{\rm jet}$=10, a
high-energy cutoff for electrons of $\gamma_{\rm emax}(z_0)=10^6$, and a small viewing
angle $\theta=1^{\circ}$ (see Table~\ref{tideal2}). This would correspond to a {\sl
microblazar}, which is also a strong non-thermal X-ray source. It is difficult to say
whether persistent microquasar jets could reach such high Lorentz factors, although it is
an interesting case to explore.

\begin{figure}
\resizebox{\hsize}{!}{\includegraphics{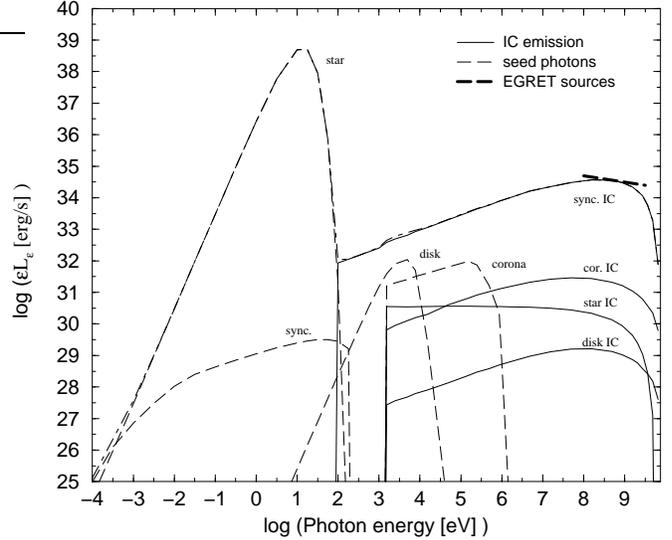}}
\caption{SED for a model with $q_{\rm e}=10^{-3}$, $B(z_0)$=200~G, $L_{\rm
cor}=3\times 10^{32}$~erg~s$^{-1}$, $\gamma_{\rm emax}(z_0)=10^4$, $\Gamma_{\rm jet}$=1.1, 
and a viewing angle of $10^{\circ}$. This is the 'realistic' case.} 
\label{realistic}
\end{figure}

Finally, we present a `realistic' case, reproducing roughly the SED observed in the two
EGRET sources that might be associated with known microquasars: LS~5039/3EG~J1824--1514
and LS I~+61 303/3EG~J0241$+$6103 (see Fig.~\ref{realistic} and Table~\ref{tideal3}). For
this particular case, we have taken both the disk and the corona to be faint, as it
appears to be the case in both sources, with $L_{\rm cor}=3\times 10^{32}$~erg~s$^{-1}$.
Also, the magnetic field required now to match the observations is $\sim 200~$G, $q_{\rm
e}$ is $\sim 10^{-3}$, and consistent with observational constraints from both
sources, a mildly relativistic jet velocity ($\Gamma_{\rm jet}=1.1$) is used. The
electron high-energy cutoff is at $\gamma_{\rm emax}(z_0)\sim 10^4$, in agreement with
EGRET data (see Fig.~\ref{realistic}). 

\begin{table}
\begin{center}
\caption[]{Range of parameters adopted for calculation of different models.}
\begin{tabular}{l c c c c c}
\noalign{\smallskip} 
\hline 
\hline \noalign{\smallskip} Parameter (symbol) & Value 
\cr \noalign{\smallskip} 
\hline \noalign{\smallskip} 
Corona luminosity ($L_{\rm cor}$) & $10^{34}$, $10^{37}$~erg~s$^{-1}$
\cr Magnetic field ($B(z_0)$) & 0.1, 100~G
\cr Bulk Lorentz factor of the jet ($\Gamma_{\rm jet}$) & 1.5, 5
\cr Disk/jet coupling constant ($q_{\rm e}$) & $10^{-5}$, $10^{-3}$
\cr \noalign{\smallskip} \hline
\end{tabular}
\label{frparam}
\end{center}
\end{table}

\begin{table}
\begin{center}
\caption[]{Parameters for a representative microquasar model.}
\begin{tabular}{l c c c c c}
\noalign{\smallskip} 
\hline 
\hline \noalign{\smallskip} Parameter (symbol) & Value 
\cr \noalign{\smallskip} 
\hline \noalign{\smallskip} 
Corona luminosity ($L_{\rm cor}$) & $10^{35}$~erg~s$^{-1}$
\cr Magnetic field ($B(z_0)$) & 10~G
\cr Maximum electron Lorentz factor ($\gamma_{\rm emax}(z_0)$) & 10$^4$
\cr Bulk Lorentz factor of the jet ($\Gamma_{\rm jet}$) & 2.5
\cr Disk/jet coupling constant ($q_{\rm e}$) & $10^{-4}$
\cr \noalign{\smallskip} \hline
\end{tabular}
\label{tideal1}
\end{center}
\end{table}

\begin{table}
\begin{center}
\caption[]{Parameters for an extreme case (microblazar).}
\begin{tabular}{l c c c c c}
\noalign{\smallskip} \hline 
\hline 
\noalign{\smallskip} Parameter (symbol) & Value 
\cr \noalign{\smallskip} \hline \noalign{\smallskip} 
Corona luminosity ($L_{\rm cor}$) & $10^{35}$~erg~s$^{-1}$
\cr Magnetic field ($B(z_0)$) & 10~G
\cr Maximum electron Lorentz factor ($\gamma_{\rm emax}(z_0)$) & 10$^6$
\cr Jet bulk Lorentz factor ($\Gamma_{\rm jet}$) & 10
\cr Viewing angle to jet's axis $\theta$ & $1^{\circ}$
\cr Disk/jet coupling constant ($q_{\rm e}$) & $10^{-4}$
\cr \noalign{\smallskip} \hline
\end{tabular}
\label{tideal2}
\end{center}
\end{table}

\begin{table}
\begin{center}
\caption[]{Parameters for a `realistic' case.}
\begin{tabular}{l c c c c c}
\noalign{\smallskip} 
\hline 
\hline \noalign{\smallskip} Parameter (symbol) & Value 
\cr \noalign{\smallskip} \hline \noalign{\smallskip} 
Corona luminosity ($L_{\rm cor}$) & 3$\times 10^{32}$~erg~s$^{-1}$
\cr Magnetic field ($B(z_0)$) & 200~G
\cr Maximum electron Lorentz factor ($\gamma_{\rm emax}(z_0)$) & 10$^4$
\cr Jet bulk Lorentz factor ($\Gamma_{\rm jet}$) & 1.1
\cr Viewing angle to jet's axis $\theta$ & $10^{\circ}$
\cr Disk/jet coupling constant ($q_{\rm e}$) & $10^{-3}$
\cr \noalign{\smallskip} \hline
\end{tabular}
\label{tideal3}
\end{center}
\end{table}

\subsection{Electron energy evolution} \label{elenev}

In this subsection, we discuss whether our parameterization is consistent with the
local cooling rates of the electrons due to energy losses. This will
allow us to know to what extent the evolution of the electrons is well described through the adopted
parameterization, that corresponds to adiabatic expansion. If radiative losses are smaller than
adiabatic ones, then the parameterization is fine at least as a first order of approximation.

\begin{figure}
\resizebox{\hsize}{!}{\includegraphics{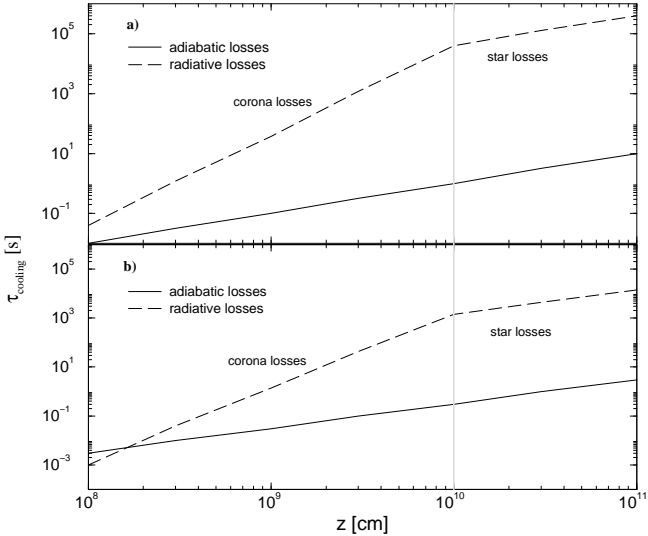}}
\caption{Cooling time evolution with $z$ for adiabatic (solid line) and radiative
(long-dashed line) losses. Two regions in the plot have been
established depending on the dominant source of radiative losses, which happen to be 
the corona IC losses close to the compact object and the star IC losses farther away.
\textbf{a)} `Realistic' case.
\textbf{b)} Representative case.} 
\label{cool}
\end{figure}

We have determined quantitatively the importance of the energy density of the different
seed photon fields, taking into account in our computations both the internal ones
(synchrotron photon and magnetic fields) and the external ones (star, disk and corona
photon fields). In Table~\ref{dens}, we show the energy densities for the mentioned
photon fields in the comoving frame at the base of the jet for the three microquasar
models we have calculated (see Tables~\ref{tideal1}, \ref{tideal2} and \ref{tideal3}). In
Fig.~\ref{cool}, we show the evolution of the cooling times along the jet for the
adiabatic and the radiative losses. This allows us to find out which one dominates at
different distances. To compute and compare the different cooling times, we have taken
the electron Lorentz factor which would correspond in our model to a given value of $z$,
considering the adiabatic evolution of electrons. In this plot, we show the behavior of
the `realistic' and the representative cases as examples; the microblazar case is not too
different.

\begin{table*}
\begin{center}
\caption[]{Energy densities at the base of the jet in units of erg~cm$^{-3}$.}
\begin{tabular}{l c c c c c c c}
\cr \noalign{\smallskip} 
\hline 
\hline 
\noalign{\smallskip} 
Model & Jet velocity & Lorentz factor & magnetic & synchrotron  & star         & disk           & corona
\cr
      &  (c)         &                & field    & photon field & photon field & photon field   & photon field
\cr 
\hline
\cr Representative & 0.917c & 2.5 & 2 & 3.2 & 220 & 2.4$\times 10^5$ & 2.2$\times 10^6$
\cr Microblazar & 0.995c & 10 & 2 & 49.8 & 1.6$\times 10^4$ & 3.7 & 3.4$\times 10^4$
\cr `Realistic' & 0.417c & 1.1 & 800 & 6.3$\times 10^4$ & 7.6 & 87.8 & 8.2$\times 10^4$
\cr \noalign{\smallskip} \hline
\end{tabular}
\label{dens}
\end{center}
\end{table*}

From Fig.~\ref{cool}, it is clear that it is enough in the `realistic' case to account only
for adiabatic losses to describe the electron energy evolution along the jet, whereas it is
not sufficient for the representative case at small values of $z$. In the second
situation, however, this is not a problem since we do not pretend to describe the
microphysics within the jet, but only to parametrize it in such a way that the expected
emission will be observed. This parameterization could take into account reacceleration
processes acting on electrons. The assumption that electrons are accelerated along the jet is
natural for many jet models (see, i.e., Biermann \& Strittmater \cite{Biermann87}). 
Future work will be done to implement this in a more explicit way.

\section{Comments} \label{disc}

We can see from Figs. \ref{fig3}-\ref{fig6} that for several models with disk/jet  coupling
constant $q_{\rm e}=10^{-3}$ we can get the expected luminosities in the observer's  frame
inferred for GRP I sources with the right photon index at energies $\sim1$ GeV,  i.e.
$\Gamma\sim 2$. When the magnetic field is strong enough ($B(z_0)\sim 100$ G),  SSC emission
alone can account for luminosities $\sim 10^{35}$ erg s$^{-1}$ at 1 GeV.  Models with bulk
Lorentz factors $\Gamma_{\rm jet}=1.5$ and $5$ do not produce  dramatically different results
for a jet with a viewing angle of $\sim 10^{\circ}$  except for the case of the scattered
stellar photons. In case of lower magnetic fields,  the IC scattering upon external  fields
clearly dominates. For instance, when $B(z_0)=0.1$ G, the IC component upon stellar  seed
photons is more than 5 orders of magnitude greater than the SSC emission at  100 MeV. For a
similar value of $q_{\rm e}$ and a $\Gamma_{\rm jet}=5$, the stellar IC component  usually
dominates over the disk and corona components. Only in the case of a powerful corona 
contribution ($L_{\rm cor}\sim 10^{37}$ erg s$^{-1}$) can the latter surpass the  IC emission
produced in the stellar field. The contribution arising from  interactions with disk photons
reaches values $\sim 10^{34}$ erg s$^{-1}$ only for  energetic jets ($q_{\rm e}=10^{-3}$).

Models with light leptonic jets ($q_{\rm e}=10^{-5}$) can produce
significant gamma-ray sources ($\sim 10^{34}$ erg s$^{-1}$ at 100~MeV) when a strong corona
is present (see Fig. \ref{fig5}). In these models, the spectrum tends to be a bit harder
than in the case of stellar photons, with our current set of assumptions (value of
$\Gamma_{\rm cor}$, etc).

Comparing Figs. \ref{ideal} and \ref{extreme} it is possible to see the differences between a
`mild' microquasar, with $\Gamma_{\rm jet}=2.5$, a viewing angle of $\theta=10^{\circ}$ and
$q=10^{-4}$, and a more `extreme' case with $\Gamma_{\rm jet}=10$ and $\theta=1^{\circ}$,
i.e. a microblazar. Since we have assumed in both cases a magnetic field of $B(z_0)=10$~G,
external IC scattering always dominates over SSC. In the `mild' microquasar the emission at
MeV-GeV energies is mainly due to upscattering of coronal photons whereas, in the extreme
case, the IC emission in the stellar field exceeds by far the other contributions. 
The hard X-ray counterpart
can be important, even beyond the cutoff for the corona
(at $\sim 150$~keV). Hence, INTEGRAL observations can be useful to unmask candidates that are
obscured at optical wavelength (see Combi et~al. \cite{Combi04} for a recent study in this
direction). 

An interesting feature of the extreme case is that the gamma-ray emission with a hard
spectrum $\Gamma\sim2$ extends up to tens of GeV. At TeV energies, the spectrum becomes
very soft due to the Klein-Nishina effect. This type of sources should be detectable with
modern Imaging Cherenkov Telescopes like HESS and MAGIC. Even systems with low-mass
companions and weak coronas might be detectable since the SSC component has luminosities
of $\sim 10^{35}$~erg~s$^{-1}$.           

In the case we have called `realistic' (Fig. \ref{realistic}), the high-energy emission
is dominated by SSC. We emphasize that this case is more realistic than the previous ones
only in the sense that it fits well the broadband spectra of known microquasars that are
suspected to be high-energy gamma-ray sources. A basic feature of these objects is their
weak X-ray luminosity, a feature that is shared by most of the unidentified EGRET
sources.

We now comment on the main differences between the models presented here and those
already published by Georganopoulos et al. (\cite{Georganopoulos02}) and Bosch-Ramon \&
Paredes (\cite{Bosch04}). The first authors consider just the interaction of a
homogeneous jet with disk and stellar photon fields. They work in the Thomson regime with
the aim of producing the hard X-ray emission observed in objects similar to Cygnus X-1 as
a result of the IC contributions. No magnetic field is considered and adiabatic losses
are not taken into account. On the contrary, we adopt a thermal Comptonization field at
hard X-rays, possibly originating in a corona around the black hole (e.g. Zdziarski et
al. \cite{Zdziarski03}), as suggested by the observation of a Compton reflection feature
and an iron K$\alpha$ line in some sources (for an alternative interpretation, see
Markoff et~al.  \cite{Markoff03b}). In this sense, our treatment is more similar to that
of Romero et al. (\cite{Romero02}). However, we include the effects of the magnetic field
as well as a more sophisticated parameterization of the jet. SSC emission, in fact, seems
to dominate for some reasonable choices of the parameters. In addition, we go to much
higher energies, taking into account the Klein-Nishina effect. Regarding the Bosch-Ramon
\& Paredes (\cite{Bosch04}) model, here we introduce several improvements, from the
corona effect up to a $z$-dependence of the magnetic field and other parameters.

\section{Discussion}

Variability in the gamma-ray emission of microquasars on timescales from days to months can
be caused by both changes in factors external to the jet or in the jet  itself. 
The winds of the companion star can change inducing variations in the
accretion rate (Reig et~al. \cite{Reig03}, McSwain et~al. \cite{McSwain04}). 
Also, if the orbit of the compact object is not completely circularized,
periodic changes in the IC flux from the stellar photon field and in the accretion rate can
be expected. These changes would vary, in turn, the power carried by the jet (through
changes in $q_{\rm e}$), the emission of the jet, and the emission of the thermal plasma around
the black hole. 
Precession of the accretion disk (with timescales of months, e.g. Brocksopp et~al.
\cite{Brocksopp99}) might also result in a time-modulation of the corresponding  soft X-ray
field. Precession of the jet itself can, additionally, produce strong variability due to the
variation of the Doppler factor (Kaufman Bernad\'o et~al.  \cite{Kaufman02}). All these
contributions might result in a complex lightcurve, with several associated timescales. 

In sources where the jet is steady and a transient feature associated with the low-hard
state, its disappearance in the high-soft state would lead, of course, to the suppression of
the gamma-ray emission. This should be an additional source of variability. This transition,
however, seems not to occur in microquasars with persistent jets like LS 5039.   

Besides medium and long-term variations, the presence of shocks in the jets can introduce
very rapid changes in the gamma-ray flux. Relativistic shocks are the natural result of
sudden changes in the injection rate of plasma in the jets. These shocks propagate
downstream increasing the energy of the particles in the fluid and amplifying the magnetic
field. When the shock finds a small feature in the underlying jet (e.g. an inhomogeneity in
the particle density or a bend in the flow direction, see Romero \cite{Romero96}) a very
rapid variation in the high-energy flux can occur. The typical timescale for these events
will be determined  by the time it takes the shock to move trough the feature. Assuming that
the feature has a size of the order of the jet radius, this gives:
\begin{equation}
t_{\rm var}\sim \frac{R_{\rm jet}}{D_{\rm s}v_{\rm s}},	
\end{equation}
where $D_{\rm s}$ is the Doppler factor of the shock and $v_{\rm s}$ its velocity. For
$R_{\rm jet}\sim 10^7$~cm, and a relativistic shock with $D_{\rm s}\sim 5$ and $v_{\rm
s}\sim c$, we have $t_{\rm var}\sim10^{-4}$~s, so this would be a very rapid flickering
superposed on the longer variability. 

Microquasars are not, of course, the only kind of galactic object that might display
variable gamma-ray emission. Other alternatives include early-type binaries (Benaglia \&
Romero \cite{Benaglia03}), accreting neutron stars (Romero et~al. \cite{Romero01b}), pulsar
wind nebulae (Roberts et~al. \cite{Roberts02}) and exotic objects (Punsly et~al.
\cite{Punsly00}). However, microquasars are perhaps the most attractive candidates to
explain a significant fraction of the variable GRP I sources because of the presence of
relativistic jets in these objects, as well as the external photon fields provided by the
companion star and the accreting matter. GRP II sources might also be associated with old
low-mass microquasars if the jets of these objects have magnetic fields strong enough to
make SSC the primary source of radiation (Kaufman Bernad\'o et~al. \cite{Kaufman04}).           

\section{Conclusion}

We have shown that the variable gamma-ray sources found on the galactic plane have some
common features that make it reasonable to consider them as a distinctive group of GRP I
sources. We have suggested that these sources might be microquasars with high-mass
stellar companions and we have developed some detailed models to explain the gamma-ray
production in this type of objects. In particular, we have considered inhomogeneous jet
models where gamma-rays are the result of inverse Compton interactions of leptons in the 
jet with locally produced synchrotron photons as well as external photon fields. We have
calculated the emission resulting from the upscattering of disk, coronal, and stellar
photons, incorporating a full Klein-Nishina calculation and the effect of losses in the
particle spectrum. We have shown that a wide variety of spectral energy distributions at
high energies can be obtained from different and reasonable combinations of the physical
parameters like magnetic field, jet power, coronal and disk luminosities, etc.  It seems
clear that the microquasar phenomenon can be naturally extended up to the highest
energies and that we can expect these objects to manifest themselves as a distinctive
group of gamma-ray sources that might be detectable with satellite-borne instruments like
those to  be carried by AGILE and GLAST, and even by ground-based Cherenkov telescopes
like HESS and MAGIC.      

\begin{acknowledgements}

We thank Marina Kaufman Bernad\'o, Felix Aharonian and Peter Biermann for useful  discussions
on microquasars. We also thank an anonymous referee for constructive comments on the
manuscript.  V.B-R. and J.M.P. acknowledge partial support by DGI of the Ministerio de
Ciencia y Tecnolog{\'{\i}}a (Spain) under grant AYA-2001-3092, as well as additional support
from the European Regional Development Fund (ERDF/FEDER). During this work, V.B-R has been
supported by the DGI of the Ministerio de Ciencia y Tecnolog{\'{\i}}a (Spain) under the
fellowship FP-2001-2699. G.E.R. has been supported by Fundaci\'on Antorchas and the Argentine
agencies ANPCyT and CONICET (PIP 0438/98). This project benefited from an international
cooperation grant funded by Fundaci\'on Antorchas. 

\end{acknowledgements}

{}


\begin{thebibliography}{}

\bibitem[1999]{Atoyan&Aharonian99}
Atoyan, A.~M.~\& Aharonian, F.~A.\ 1999, \mnras, 302, 253 

\bibitem[2003]{Benaglia03}
Benaglia, P., Romero, G.E. 2003, A\&A 399, 1121

\bibitem[2003]{Bhatta03} 
Bhattacharya, D., Aky\"uz, A. Miyagi, T., Samimi, J. ~\& Zych, A. 2003, A\&A 404, 163 

\bibitem[1987]{Biermann87} 
Biermann, P. L., Strittmatter, P. A. 1987, \apj, 322, 643 

\bibitem[1970]{Blumenthal&Gould70}
Blumenthal, G. R.~\& Gould, R. J. 1970, Rev. Mod. Phys., 42, 237

\bibitem[2004]{Bosch04}
Bosch-Ramon, V.~\& Paredes, J.~M.\ 2004, \aap, 417, 1075 

\bibitem[1999]{Brocksopp99}
Brocksopp, C., Fender, R.P., Larimov, V., et~al. 1999, \mnras, 309, 1063 

\bibitem[2004]{Combi04}
Combi, J.~A., Rib\'o, M., Mirabel, I.~F., \& Sugizaki, M.
2004, A\&A, 422, 1031

\bibitem[1992]{Dermer92}
Dermer, C.~D., Schlickeiser, R., \& Mastichiadis, A.\ 1992, \aap, 256, L27

\bibitem[2002]{Dermer&Sch02} Dermer, 
C.~D.~\& Schlickeiser, R.\ 2002, \apj, 575, 667 

\bibitem[1995]{Falcke&Biermann95}
Falcke, H.~\& Biermann, P.~L.\ 1995, \aap, 293, 665

\bibitem[1999]{Falcke&Biermann99}
Falcke, H.~\& Biermann, P.~L.\ 1999, \aap, 342, 49

\bibitem[2000]{Gehrels00} 
Gehrels, N., Macomb, D.J., Bertsch, D.L., Thompson, D. J., \& Hartman R.C. 2000, Nat 404, 363

\bibitem[2002]{Georganopoulos02}
Georganopoulos, M., Aharonian, F.~A., \& Kirk, J.~G. 2002, A\&A, 388, L25

\bibitem[1985]{Ghisellini85}
Ghisellini, G., Maraschi, L., \& Treves, A. 1985, A\&A, 146, 204 

\bibitem[1995]{Grenier95} 
Grenier, I.A. 1995, Adv. Space Res., 15, 573
 
\bibitem[2001]{Grenier01} 
Grenier, I.A. 2001, in: The Nature of Unindentified Galactic High-Energy
Gamma-Ray Sources, ed. A. Carraminana, O. Reimer, \& D. Thompson, Kluwer
Academic Publishers, Dordrecht, p.51

\bibitem[2004]{Grenier04}
Grenier, I.A. 2004, in: Cosmic Gamma-Ray Sources, ed. K.S. Cheng \& G.E. Romero, Kluwer
Academic Publishers, Dordrecht, in press

\bibitem[2002]{Grimm02} 
Grimm, H. J., Gilfanov, M., \& Sunyaev, R. 2002, A\&A, 391, 923 

\bibitem[1999]{Hartman99}
Hartman, R. C., Bertsch, D.~L., \& Bloom, S.~D.~et~al. 1999, \apjs, 123, 79

\bibitem[2002]{Kaufman02}
Kaufman Bernad\'o, M.~M., Romero, G.~E., \& Mirabel, I.~F. 2002, A\&A, 385,
L10

\bibitem[2004]{Kaufman04}
Kaufman Bernad\'o, M.~M. et al. 2004, in preparation

\bibitem[2002]{Maccarone02}
Maccarone, T.J. 2002, \mnras, 336, 1371

\bibitem[2001]{Markoff01}
Markoff, S., Falcke, H., \& Fender, R.\ 2001, \aap, 372, L25 

\bibitem[2003]{Markoff03}
Markoff, S., Nowak, M., Corbel, S., et~al. 2003, A\&A 397, 645 

\bibitem[2003b]{Markoff03b}
Markoff, S., Nowak, M., Corbel, S., et~al. 2003b, New Astronomy Review, 47, 491

\bibitem[2004]{Massi04}
Massi, M., Rib\'o, M., Paredes, J. M, et al. 2004, A\&A 414, L1

\bibitem[2004]{McClin04}
McClintock, J. E.~\& Remillard, R. A. 2004, to appear as Chapter 4 in "Compact Stellar 
X-ray Sources," eds. W.H.G. Lewin and M. van der Klis, Cambridge University Press, 
[Astro-ph/0306213]

\bibitem[2004]{McSwain04}
McSwain, M.~V., Gies, D.~R., Huang, W., et~al. 2004, \apj, 600, 927

\bibitem[1999]{Mirabel&Rodriguez99}
Mirabel, I.~F.~\& Rodr{\'{\i}}guez, L.~F. 1999, ARA\&A, 37, 409

\bibitem[2004]{Mirabel04}
Mirabel, I. F., Rodrigues, I., \& Liu, Q. Z. 2004 A\&A, 422, L29

\bibitem[2004]{Miyagi04}
Miyagi, T. \& Bhattacharya, D. 2004, personal communication

\bibitem[2003]{Nolan03}
Nolan, P.L, Tompkins, W.F., Grenier, I.A., Michelson, P.F. 2003, ApJ, 597, 615 

\bibitem[1970]{Pacholczyk70}
Pacholczyk, A.~G., 1970, Radio Astrophysics, Freeman, San Francisco, CA

\bibitem[2000]{Paredes00}
Paredes, J.M., Mart\'{\i}, J., Rib\'{o}, M.,  Massi, M. 2000, Science, 288, 2340

\bibitem[2002]{Paredes02}
Paredes, J. M., Rib\'o, M., Ros, E., et al. 2002, A\&A 393, L99

\bibitem[1998]{Poutanen98}
Poutanen, J. 1998, in: Theory of Black Hole Accretion Disks, M. Abramowicz et~al. (eds), Cambridge University Press, Cambridge, p. 100

\bibitem[2000]{Punsly00}
Punsly, B., Romero, G.E., Torres, D. F., Combi, J. A. 2000, A\&A, 364, 552

\bibitem[2003]{Reig03}
Reig, P., Rib\'o, M., Paredes, J.~M., \& Mart\'{\i}, J. 2003, A\&A, 405, 285

\bibitem[2001]{Reimer01} 
Reimer, O. 2001, in: The Nature of Unindentified Galactic High-Energy
Gamma-Ray Sources, ed. A. Carraminana, O. Reimer, \& D. Thompson, Kluwer
Academic Publishers, Dordrecht, p.17

\bibitem[2002]{Ribo02}
Rib\'o M., Paredes J.M., Romero G.E., et~al., 2002, A\&A, 384, 954

\bibitem[2002]{Roberts02}
Roberts, M.S.E.; Gaensler, B.M., Romani, R.W. 2002,
in: Neutron Stars in Supernova Remnants, ASP Conference Series, Vol. 271, Patrick O. Slane and Bryan M. Gaensler (eds.), San Francisco: ASP, 2002., p.213

\bibitem[1996]{Romero96}
Romero, G.~E. 1996, A\&A, 313, 759

\bibitem[1999]{Romero99}
Romero, G.~E., Benaglia, P., Torres, D.~F. 1999, A\&A, 348, 868

\bibitem[2001]{Romero01}
Romero, G.~E. 2001, in: The Nature of Unindentified Galactic
High-Energy Gamma-Ray Sources, ed. A. Carraminana, O. Reimer, \& D. Thompson, Kluwer
Academic Publishers, Dordrecht, 65

\bibitem[2001]{Romero01b}
Romero, G.E., Kaufman Bernad\'o, M.M., Combi, J.A., Torres, D.F. 2001, A\&A, 376, 599

\bibitem[2002]{Romero02}
Romero, G.E., Kaufman Bernad\'o, M.M., \& Mirabel, I.F. 2002,
A\&A, 393, L61

\bibitem[2003]{Romero03}
Romero, G.~E., Torres, D.~F., Kaufman  Bernad\'o, M.~M., \& Mirabel, I.~F. 2003, A\&A, 410, L1  

\bibitem[2004]{Romero04}
Romero, G.~E., Grenier, I.~A., Kaufman Bernad\'o, M.M., \& Mirabel, I.F., \& Torres, D.~F. 2004, ESA-SP, in press [astro-ph/0402285]

\bibitem[2001]{Torres01} 
Torres, D.~F., Romero, G.~E., Combi, J.~A., Benaglia, P., Andernach, H., \& Punsly, B.\ 2001, 
\aap, 370, 468 

\bibitem[2003]{Zdziarski03}
Zdziarski, A.~A., Lubinski, P., Gilfanov, M., Revnivtsev, M. 2003, MNRAS 342, 355

\end{thebibliography}
\end{document}